\documentclass[aps,pre,reprint,showpacs]{revtex4-1}

\usepackage{graphicx}
\usepackage{amsmath}
\usepackage{amssymb}
\usepackage[usenames, dvipsnames]{color}
\usepackage[normalem]{ulem}
\begin{document}

\preprint{}

\title{The geometry of thermodynamic control}

\author{Patrick R. Zulkowski}
\email[]{pzulkowski@berkeley.edu}
\affiliation{Department of Physics, University of California, Berkeley, CA 94720}
\affiliation{Redwood Center for Theoretical Neuroscience, University of California, Berkeley CA 94720}

\author{David A. Sivak}
\email[]{dasivak@lbl.gov}
\altaffiliation{Present address: Center for Systems and Synthetic Biology, University of California, San Francisco, CA 94158; david.sivak@ucsf.edu}
\affiliation{Physical Biosciences Division, Lawrence Berkeley National Laboratory, Berkeley, CA 94720}

\author{Gavin E. Crooks}
\email[]{gcrooks@lbl.gov}
\affiliation{Physical Biosciences Division, Lawrence Berkeley National Laboratory, Berkeley, CA 94720}

\author{Michael R. DeWeese}
\email[]{deweese@berkeley.edu}
\affiliation{Department of Physics, University of California, Berkeley, CA 94720}
\affiliation{Redwood Center for Theoretical Neuroscience, University of California, Berkeley CA 94720}
\affiliation{Helen Wills Neuroscience Institute, University of California, Berkeley CA 94720}

\begin{abstract}
A deeper understanding of nonequilibrium phenomena is needed to reveal the principles governing natural and synthetic molecular machines. Recent work has shown that when a thermodynamic system is driven from equilibrium then, in the linear response regime, the space of controllable parameters has a Riemannian geometry induced by a generalized friction tensor. We exploit this geometric insight to construct closed-form expressions for minimal-dissipation protocols for a particle diffusing in a one dimensional harmonic potential, where the spring constant, inverse temperature, and trap location are adjusted simultaneously. These optimal protocols are geodesics on the Riemannian manifold, and reveal that this simple model has a surprisingly rich geometry. We test these optimal protocols via a numerical implementation of the Fokker-Planck equation and demonstrate that the friction tensor arises naturally from a first order expansion in temporal derivatives of the control parameters, without appealing directly to linear response theory.
\end{abstract}

\pacs{05.70.Ln,  02.40.-k,05.40.-a}

\date{\today}

\maketitle

\section{Introduction}

There has been considerable progress in the study of nonequilibrium processes in recent years. For example, fluctuation theorems relating the probability of an increase to that of a comparable decrease in entropy during a finite time period have been derived~\cite{Evans1993,Evans1994,Gallavotti1995a,Crooks1999,Hatano2001} and experimentally verified~\cite{Wang2002,Carberry2004,Garnier2005,Toyabe2010} in a variety of contexts. Moreover, other new fundamental relationships between thermodynamic quantities that remain valid even for systems driven far from equilibrium, such as the Jarzynski equality~\cite{Jarzynski1997,Liphardt2002,Seifert2005b,Sagawa2010}, have also been established. Interestingly, some of these ideas were independently developed in parallel within the machine learning community~\cite{Neal2001}, as ideas from nonequilibrium statistical mechanics are increasingly finding applications to learning and inference problems~\cite{Nilmeier2011,Still2012}.

For macroscopic systems, the properties of optimal driving processes have been investigated using thermodynamic length, a natural measure of the distance between pairs of equilibrium thermodynamics states \cite{Weinhold1975a,Ruppeiner1979,Schlogl1985,Salamon1984,Salamon1983a,Brody1995}, with extensions to microscopic systems involving a metric of Fisher information \cite{Crooks2007c,Burbea1982}. Recently, a linear-response framework has been proposed for protocols that minimize the dissipation during nonequilibrium perturbations of microscopic systems. In the resulting geometric formulation, a generalized friction tensor induces a Riemannian manifold structure on the space of parameters, and optimal protocols trace out geodesics of this friction tensor~\cite{Sivak2012b}.

In this article, we make use of Riemannian geometry theorems to simplify the problem of optimizing protocols. To illustrate the power of these geometric ideas, we consider a simple, but previously unsolved, stochastic system and calculate closed-form expressions for optimal protocols. We test the accuracy of our approximation by numerically comparing our optimal protocols against naive protocols using the Fokker-Planck equation. We conclude by demonstrating that our inverse diffusion tensor framework arises naturally from a first order expansion in temporal derivatives of the control parameters, without appealing directly to linear response theory.\\

%\section{Geometric framework for finding optimal protocols}
\section{Derivation of the excess power for variable temperature}
For a physical system at equilibrium in contact with a thermal bath, the probability distribution over microstates $ x $ is given by the canonical ensemble
\begin{equation}\label{canonical} \pi( x | \lambda) \equiv \exp{ \beta \left[ F(\boldsymbol \lambda) - E(x,\boldsymbol \lambda) \right] } \ ,  \end{equation}
where $ \beta = (k_{\rm B} T)^{-1} $ is the inverse temperature in natural units, $ F(\boldsymbol \lambda) $ is the free energy, and $ E(x,\boldsymbol \lambda) $ is the system energy as a function of the microstate $ x $ and a collection of experimentally controllable parameters $ \boldsymbol \lambda $.

In equilibrium, the thermodynamic state of the system (the probability distribution over microstates) is completely specified by values of the control parameters, but out of equilibrium the system's probability distribution over microstates fundamentally depends on the history of the control parameters $\boldsymbol\lambda $, which we denote by the control parameter protocol $\boldsymbol \Lambda$. We assume the protocol to be sufficiently smooth to be twice-differentiable.

%Because we are considering a system with variable temperature, we need to modify
The usual expressions for heat and work~\cite{Crooks1998,Sekimoto1998,Jarzynski1998,Imparato2007} assume that the temperature of the heat bath is held constant over the course of the nonequilibrium protocol. Following the development of methods to handle time-varying temperature described in section 1.5 of~\cite{Crooks_PhD_thesis}, and preceding Eq.~(4) of \cite{Jarzynski_1999}, we argue that the unitless energy $\beta E(x,{\boldsymbol \lambda})$ (normalized by the natural scale of equilibrium thermal fluctuations, $k_{\rm B}T = \beta^{-1}$, set by equipartition) is the fundamental thermodynamic quantity. Thus when generalizing to a variable heat bath temperature, we arrive at the following definition for the average instantaneous rate of (unitless) energy flow into the system:
\begin{equation}\label{natenergy} \left< \frac{d}{dt} \bigg( \beta E(x, \boldsymbol \lambda) \bigg) \right>_{\boldsymbol \Lambda} \ , \end{equation}
where angled brackets with subscript indicate a nonequilibrium average dependent on the protocol $\boldsymbol \Lambda$. For constant $\beta$, this reduces to the standard thermodynamic definition~\cite{Sivak2012b}.
With this definition, we can prove that for systems obeying Fokker-Planck dynamics, excess work is guaranteed to be non-negative for any path, which is not true of the naive definition (see \textsection~\ref{model_system}). Nonetheless, a deeper understanding of the subtleties involved in our modified energy flow definition (Eq.~(\ref{natenergy})) calls out for further study.
%~\ref{model_system}).}
%\prz{[PRZ: anomalies in the computed excess work values assuming that $ \langle W_{\rm ex} \rangle \equiv \int_{t_{0}}^{t_{f}} \mathcal{P}_{\rm ex} dt $; i.e. that the appropriate definition of excess work involved excess power without the factor of $ \beta $.
%David and I looked into this more deeply and found that defining $ \langle \beta W_{\rm ex} \rangle \equiv \int_{t_{0}}^{t_{f}} \beta \mathcal{P}_{\rm ex} dt $ remedied the problem, at least for all protocols I used in my numerical solver.
%David argued (and David can correct me on this if I am mistaken) that $ \beta $ sets the natural scale for energy fluctuations of the system. Therefore, to properly compare energies at different times (so as to properly integrate power over time and obtain excess work, for instance), one must rescale at each time $ t $ by $ \beta(t)$. Moreover, it is mathematically possible (between the write-up I sent to you a few days ago and the Jarzynski paper that David sent around) to show that the excess work $ \langle \beta W_{\rm ex} \rangle \equiv \int_{t_{0}}^{t_{f}} \beta \mathcal{P}_{\rm ex} dt $ must be strictly positive if our system begins in equilibrium. ]}

Eq.~\eqref{natenergy} may be written as
\begin{equation} \left< \frac{d x^{T}}{dt} \cdot  \frac{\partial \left( \beta E \right)}{\partial x}(x, \boldsymbol \lambda) \right>_{\boldsymbol \Lambda} + \left< \frac{d \boldsymbol \lambda^{T}}{dt} \cdot  \frac{\partial \left( \beta E \right)}{\partial \boldsymbol \lambda}(x, \boldsymbol \lambda) \right>_{\boldsymbol \Lambda}. \end{equation}
The first term represents energy flux due to fluctuations of the system at constant parameter values and naturally defines heat flux for nonequilibrium systems. The second term, associated with an energy flux due to changes of the external parameters, defines nonequilibrium average power in the general setting of time-variable bath temperature.

The average excess power exerted by the external agent on the system, over and above the average power on a system at equilibrium, is
\begin{equation}\label{expower} \beta(t_{0}) \mathcal{P}_{\rm ex}(t_{0}) \equiv - \left[ \frac{ d \boldsymbol \lambda^{T} }{d t} \right]_{t_{0}} \cdot \left< \triangle \boldsymbol X \right>_{ \boldsymbol \Lambda} \ . \end{equation}
Here $ \boldsymbol X \equiv -\frac{ \partial \left( \beta E \right)}{\partial \boldsymbol \lambda} $ are the forces conjugate to the control parameters $ \boldsymbol \lambda $, and $ \triangle \boldsymbol X (t_{0}) \equiv \boldsymbol X(t_{0}) - \left< \boldsymbol X \right>_{\boldsymbol \lambda(t_{0}) } $ is the deviation of ${\bf X}(t_0)$ from its current equilibrium value.

%Applying linear response theory~\cite{Zwanzig2001} for protocols that vary sufficiently slowly~\cite{Sivak2012b}, the resulting mean excess power is
%\begin{equation}\label{etadef} \beta(t_{0}) \mathcal{P}_{\rm ex}(t_{0}) \approx  \left[ \frac{ d \boldsymbol \lambda^{T} }{d t}  \right]_{t_{0}} \cdot g(\boldsymbol \lambda(t_{0}) ) \cdot \left[ \frac{ d \boldsymbol \lambda }{d t}  \right]_{t_{0}} \ , \end{equation}
%for inverse diffusion tensor
%\begin{equation}\label{newfrictensor} g_{ij} \equiv \beta(t_{0}) \zeta_{ij}(\boldsymbol \lambda(t_{0})) =  \int_{0}^{\infty} dt' \big\langle \delta X_{j}(0) \delta X_{i}(t') \big\rangle_{\boldsymbol \lambda(t_{0})}\ , \end{equation}
%where $ \zeta_{ij} $ is the friction tensor in control parameter space from~\cite{Sivak2012b}. We will construct geodesics using this inverse diffusion tensor $g_{ij}$. \\

Applying linear response theory~\cite{Zwanzig2001},
\begin{equation} \left< \triangle \boldsymbol  X(t_{0}) \right>_{ \boldsymbol \Lambda} \approx \int_{-\infty}^{t_{0}} \boldsymbol \chi(t_{0}-t') \cdot \left[ \boldsymbol \lambda(t_{0}) - \boldsymbol \lambda(t') \right] dt' \end{equation}
where $ \chi_{ij}(t) \equiv - d\Sigma_{ij}^{\boldsymbol \lambda(t_{0})}(t)/dt $ represents the response
of conjugate force $ X_{i} $ at time $ t $ to a perturbation in control
parameter $ \lambda^{j} $ at time zero, and
\begin{equation} \Sigma_{ij}^{\boldsymbol \lambda(t_{0})}(t) \equiv \left< \delta X_{j}(0) \delta X_{i}(t) \right>_{\boldsymbol \lambda(t_{0})}. \end{equation}
For protocols that vary sufficiently slowly~\cite{Sivak2012b}, the resulting mean excess power is
\begin{equation}\label{etadef} \beta(t_{0}) \mathcal{P}_{\rm ex}(t_{0}) \approx  \left[ \frac{ d \boldsymbol \lambda^{T} }{d t}  \right]_{t_{0}} \cdot g(\boldsymbol \lambda(t_{0}) ) \cdot \left[ \frac{ d \boldsymbol \lambda }{d t}  \right]_{t_{0}} \ , \end{equation}
for inverse diffusion tensor
\begin{equation}\label{newfrictensor} g_{ij} \equiv \beta(t_{0}) \zeta_{ij}(\boldsymbol \lambda(t_{0})) =  \int_{0}^{\infty} dt' \big\langle \delta X_{j}(0) \delta X_{i}(t') \big\rangle_{\boldsymbol \lambda(t_{0})}\ , \end{equation}
where $\zeta_{ij}$ is the friction tensor in control parameter space from~\cite{Sivak2012b}. We will construct geodesics using this inverse diffusion tensor $g_{ij}$. \\

\section{The model system and its inverse diffusion tensor \label{model_system}}

\begin{figure}
\begin{center}
(a)
\\ \includegraphics{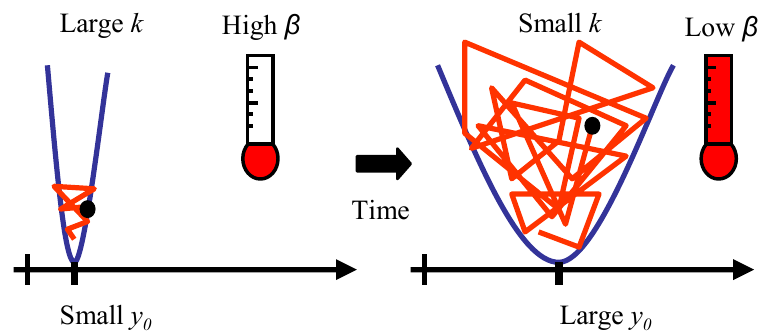}
\\~\\ (b) \hspace{10em} (c)
\\
\includegraphics{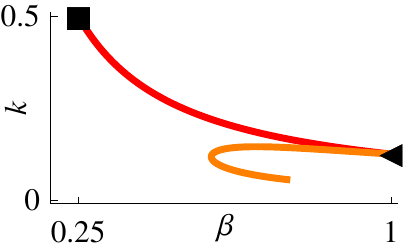}
\includegraphics{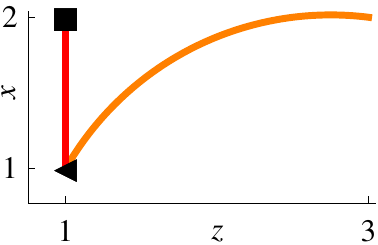}
\\~\\
(d)
\\ \includegraphics{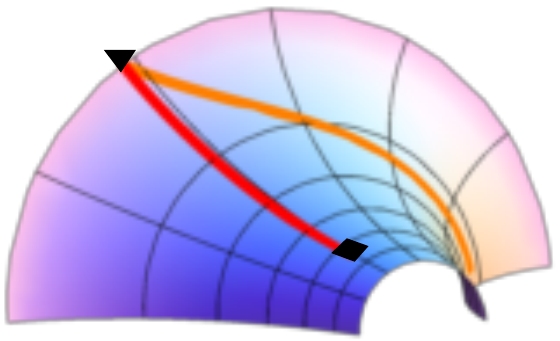}
\end{center}
\caption{\textbf{(a)} Our model system. A particle (black dot) diffusing in a harmonic potential with adjustable spring constant $ k $, position $ y_{0} $, and inverse temperature $ \beta = \frac{1}{k_{\rm B} T} $ (indicated by thermometers).
\textbf{(b)}
Representative optimal protocols (orange and red curves) plotted for two of the three control parameters, $k$ and $\beta$. An optimal protocol ({\em e.g.}, red curve) results in the minimum dissipation for any path taking the system from one particular state (black square) to another (black triangle) in a fixed amount of time.
\textbf{(c)}
A change of variables $\{\beta,k\}\rightarrow \{z,x\}$ (Eq.~(\ref{confcoord})) reveals that our model system has an underlying structure described by hyperbolic geometry, represented here as the Poincar\'{e} half-plane, in which geodesics form half circles (orange curve) or vertical lines (red line).
\textbf{(d)} A piece of the $(z,x)$ manifold may be isometrically embedded as a saddle in $ \mathbb{R}^3 $. The distortions in each of these two optimal paths as shown in panels (b) and (c) reflect the curvature of this manifold.}
\label{fig:geofig}
\end{figure}

We consider a particle (initially at equilibrium) in a one-dimensional harmonic potential diffusing under inertial Langevin dynamics, with equation of motion
\begin{equation} m \ddot{y} + k(t) \left( y - y_{0}(t) \right) + \zeta^c \dot{y} = F(t) \ , \end{equation}
for Gaussian white noise $F(t)$ satisfying
\begin{equation} \left\langle F(t) \right\rangle = 0 \ , \quad \left< F(t) F(t') \right> = \frac{2 \zeta^c}{\beta(t)} \delta (t-t') \ . \end{equation}
Here $\zeta^c$ is the Cartesian friction coefficient. We take as our three control parameters: the inverse temperature of the bath $\beta$, the location of the harmonic potential minimum $y_{0}$, and the stiffness of the trap $k$ [see Fig.~\ref{fig:geofig}(a)]. The conjugate forces are
\begin{equation}
\boldsymbol X = \left( \beta k \left( y- y_{0} \right), - \frac{p^2}{2m} - \frac{k}{2} \left( y-y_{0} \right)^2, - \frac{\beta}{2} \left( y-y_{0} \right)^2 \right).
\end{equation}
This model can be experimentally realized as, for instance, a driven torsion pendulum~\cite{Douarche2005a,Ciliberto2010}.

The excess work
\begin{equation}
\langle (\beta W)_{\rm ex} \rangle \equiv \int_{t_a}^{t_b} dt \ \beta(t) \mathcal{P}_{\rm ex}(t)
\end{equation}
is non-negative.
%A proof of this statement is available when the Fokker-Planck equation Eq.~\eqref{FokkerPlanck} determines the nonequilibrium probability distribution $f(y,p,t) \equiv f$.
Assuming the system begins in equilibrium, the relative entropy $D\big[ f \ || \ \pi(x| \lambda(t)) \big]$ corresponds to the available energy in the system due to being out of equilibrium~\cite{Sivak2012a}, and bounds the excess work from below. Here, $\pi$ is the equilibrium distribution (Eq.~\eqref{canonical}) defined by parameters $\boldsymbol \lambda(t)$, and $f \equiv f(y,p,t)$ is the nonequilibrium probability distribution. The time derivative of the relative entropy may be written as
\begin{equation}\label{KLderiv} \frac{d}{dt}D\big[ f \ || \ \pi(x| \lambda(t)) \big] = \int \frac{\partial f}{\partial t} \log \left( f /\pi(x| \lambda(t))  \right) + \beta(t) \mathcal{P}_{ex}(t),  \end{equation}
which follows from the identity
\begin{equation} \int f \left[\frac{\partial}{\partial t} \log \left( \pi(x| \lambda(t))\right) \right] = \frac{d \boldsymbol \lambda^{T} }{dt} \cdot \left< \triangle \boldsymbol X (t) \right>_{\boldsymbol \Lambda}. \end{equation}
The first term of Eq.~\eqref{KLderiv} simplifies to
\begin{equation} -\frac{\zeta^c}{\beta} \int e^{-\frac{ \beta p^2}{m}} \left[\frac{1}{f} \left( \frac{\partial}{\partial p} \left( e^{\frac{ \beta p^2}{2 m}} f \right) \right)^2 \right] \leq 0. \end{equation}
Integrating Eq.~\eqref{KLderiv} from $0$ to $t_{0}$ proves the relative entropy bounds the excess work from below. Since this quantity is always  non-negative, so is the excess work; in fact, for any finite-duration path visiting more than one point in parameter space, it is strictly positive, yielding a well-behaved metric in our geometrical formalism. See~\cite{Vaikuntanathan2009} for related results
in the special case of constant temperature. Note that, unlike our modified definition for work, the naive definition $\int_{t_a}^{t_b} dt \, \mathcal{P}_{\rm ex}(t)$ may be negative for particular protocols that vary $\beta$.

Calculation of the time correlation functions in Eq.~\eqref{newfrictensor} requires knowledge of the dynamics for fixed control parameters. We may write any solution to the equation of motion as a sum~$y_{h} + y_{p}$ of a homogeneous part $y_{h}$, which depends on the initial conditions and is independent of~$F(t)$, and a particular part~$y_{p}$, which has vanishing initial conditions but depends linearly on $ F(t) $ (see, for instance, Theorem 3.7.1 in~\cite{Boyce}). Explicitly, we may write
\begin{equation}\label{eq:particular} y_{p}(t) = \int_{0}^{t} \left( \frac{ y^{(1)}_{h}(s) y^{(2)}_{h}(t)-y^{(1)}_{h}(t) y^{(2)}_{h}(s) }{y^{(1)}_{h}(s) \frac{ d }{ds} y^{(2)}_{h}(s)-y^{(2)}_{h}(s) \frac{d}{ds}y^{(1)}_{h}(s)} \right) \frac{F(s)}{m}  ds  \end{equation}
where $y^{(i)}_{h}(t)$ for $i = 1,2$ are independent solutions of the homogeneous equation. It follows immediately that
\begin{equation} \label{eq:homogeneous} y_{h}(t) = C_{1} y^{(1)}_{h}(t) + C_{2} y^{(2)}_{h}(t) \end{equation}
where the constants $C_{1}, C_{2}$ are determined by initial conditions.

For Gaussian white noise $F(t)$, it is easy to show that the particular piece $y_{p}$ does not contribute to the equilibrium time correlation function $\left< \delta X_{j}(0) \delta X_{i}(t) \right>$. For simplicity and without loss of generality, consider the correlation function $\left< \delta y(t)^2 \delta y(0)^2 \right>$. Expanding this expression,
\begin{equation}
\left< \delta y(t)^2 \delta y(0)^2 \right> = \left< y(t)^2  y(0)^2 \right> - \left< y(t)^2 \right> \left< y(0)^2 \right> \ ,
\end{equation}
and substituting $y(t) = y_{h}(t)+y_{p}(t)$, we find
\begin{align}
\big< \delta y(t)^2 &\delta y(0)^2 \big> = \left< y_{h}(t)^2  y(0)^2 \right> + \left< y_{p}(t)^2  y(0)^2 \right> \nonumber\\
&- \left< y_{h}(t)^2 \right> \left< y(0)^2\right> - \left< y_{p}(t)^2 \right> \left< y(0)^2 \right> \\
&+ 2 \Big( \left< y_{h}(t) y_{p}(t)  y(0)^2 \right> - \left< y_{h}(t) y_{p}(t) \right> \left< y(0)^2 \right> \Big) \ . \nonumber
\end{align}
Angled brackets denote an average over noise and initial conditions.

According to Eq.~\eqref{eq:particular}, the particular solution $y_{p}$ does not depend on the initial conditions. It follows immediately that
\begin{equation} \left< y_{p}(t)^2  y(0)^2 \right> - \left< y_{p}(t)^2 \right> \left< y(0)^2 \right> = 0 \ . \end{equation}
Furthermore, since $y_{h}$ depends only on the initial conditions and is independent of the noise,
\begin{equation} \left< y_{h}(t) y_{p}(t)  y(0)^2 \right> - \left< y_{h}(t) y_{p}(t) \right> \left< y(0)^2 \right> = 0 \ , \end{equation}
follows from the assumption that $\left< F(t) \right> = 0$. To summarize,
\begin{equation} \left< \delta y(t)^2 \delta y(0)^2 \right> = \left< \delta y_{h}(t)^2 \delta y(0)^2 \right>. \end{equation} For each of the time correlation functions needed to compute the inverse diffusion tensor, it is generally true that $y_{h}(t)$ may be substituted in the average for $y(t)$.

Without loss of generality, let us assume for the moment that $\left( \zeta^{c} \right)^2-4 k m > 0$. If we define
\begin{equation}  r_{\pm} = \frac{\zeta^{c}}{2m} \pm \frac{1}{2} \sqrt{ \left( \frac{\zeta^{c}}{m} \right)^2 - \frac{ 4 k}{m} } \ , \end{equation}
then the homogeneous solution with initial conditions $\{y(0), p(0) = m\dot{y}(0)\}$ is given by
\begin{eqnarray} y_{h}(t) & = & y_{0}+ \frac{p(0)+ m r_{-} \left(y(0)-y_{0} \right) }{m (r_{-}-r_{+})}e^{-r_{+}t} \nonumber \\ && + \frac{p(0)+ m r_{+}  \left(y(0)-y_{0} \right) }{m (r_{+}-r_{-})}e^{-r_{-}t} \ , \end{eqnarray}
where~$y_{0}$ is the fixed trap position. For convenience, let us define $Y \equiv y-y_{0}$. Assuming that the initial conditions $\{y(0),p(0)\}$ are distributed according to the equilibrium Boltzmann distribution~$\pi[y(0),p(0)] \propto e^{-\beta E[y(0),p(0)]}$ for $E[y,p] = \frac{p^2}{2 m}  + \frac{1}{2} k Y^2$, we obtain the following identities:
\begin{subequations}
\begin{align}
 \langle \delta Y_{h}^{2}(t) \delta Y^{2}(0) \rangle & = \frac{2}{\left( k \beta \right)^2 \left( r_{+}-r_{-} \right)^2 } \left( r_{-} e^{-r_{+} t} - r_{+} e^{-r_{-} t} \right)^2 \\
  \langle \delta \dot{Y}_{h}^{2}(t) \delta Y^{2}(0) \rangle & = \frac{2 r_{+}^2 r_{-}^2}{k^2 \beta^2 \left( r_{+}-r_{-} \right)^2} \left( e^{-r_{+} t } - e^{-r_{-} t} \right)^2 \\
 \langle \delta Y_{h}^{2}(t) \delta p^{2}(0) \rangle & = \frac{2}{\beta^2 \left( r_{+}-r_{-} \right)^2} \left( e^{-r_{+} t } - e^{-r_{-} t} \right)^2 \\
 \langle \delta \dot{Y}_{h}^{2}(t) \delta p^{2}(0) \rangle &= \frac{2}{\beta^2 \left( r_{+}-r_{-} \right)^2} \left( r_{-} e^{-r_{-} t} - r_{+} e^{-r_{+} t} \right)^2.
\end{align}
\end{subequations}
Integrating these expressions, we obtain
\begin{subequations}
\begin{align}
\int_{0}^{\infty} dt \langle \delta Y_{h}^{2}(t) \delta Y^{2}(0) \rangle &= \frac{m}{k^2 \beta^2 \zeta^{c}} \left( 1 + \frac{\left( \zeta^{c} \right)^2}{k m} \right) \\
\int_{0}^{\infty} dt \langle \delta \dot{Y}_{h}^{2}(t) \delta Y^{2}(0) \rangle &= \frac{1}{k \beta^2 \zeta^{c}} \\
\int_{0}^{\infty} dt \langle \delta Y_{h}^{2}(t) \delta p^{2}(0) \rangle &= \frac{m^{2}}{k \beta^{2} \zeta^{c}} \\
\int_{0}^{\infty} dt \langle \delta \dot{Y}_{h}^{2}(t) \delta p^{2}(0) \rangle &= \frac{m}{\zeta^{c} \beta^{2}} \ .
\end{align}
\end{subequations}

Thus the inverse diffusion tensor is
\begin{equation}\label{betafric} g_{ij} = \frac{m}{4 \zeta^{c}} \left(
     \begin{array}{ccc}
       \frac{4 \left( \zeta^{c} \right)^2}{m} \beta &  0 & 0 \\
       0 & \frac{1}{ \beta^2 } \left(4 + \frac{\left( \zeta^{c}\right)^2}{k m}  \right) & \frac{1}{ \beta k } \left( 2 + \frac{\left( \zeta^{c}\right)^2}{k m} \right) \\
        0 & \frac{1}{ \beta k } \left( 2 + \frac{\left( \zeta^{c}\right)^2}{k m} \right) & \frac{1}{ k^2  } \left( 1+ \frac{\left( \zeta^{c}\right)^2}{k m} \right) \\
     \end{array}
   \right) \ , \end{equation}
which endows the space $ -\infty < y_{0} < \infty , 0 < \beta < \infty  , 0< k < \infty  $ with a Riemannian structure.\\

\section{Brief review of Riemannian geometry}
We recall some definitions from Riemannian geometry and establish notation (see~\cite{doCarmo,Jost,Carroll} for details). For a smooth Riemannian manifold $M$ endowed with metric tensor $g$, %the covariant derivative of a smooth vector field $ V $ is given in a local coordinate system by
%\begin{equation} \nabla_{i} V^{j} \equiv \partial_{i} V^{j}+ \Gamma^{j}_{ik} V^{k}, \end{equation}
the Christoffel symbols are defined as
\begin{equation}\label{Christoffel} \Gamma^{j}_{ik} \equiv \frac{1}{2} g^{jl} \left( \partial_{i} g_{kl} + \partial_{k} g_{il}-\partial_{l} g_{ik} \right) \end{equation}
where $ g^{ij} $ denotes the matrix inverse of the metric. We employ the Einstein summation convention here (and assume it throughout). % Vector fields such as $ V $ have upper indices which may be lowered by using the metric; i.e. $ V_{i}= g_{ij} V^{j} $.
The Riemann tensor, constructed from the Christoffel symbols, measures the curvature of the manifold $ M $ and is given in local coordinates by
\begin{equation}\label{Riemann} R^{i}{}_{jkl} \equiv \partial_{k} \Gamma^{i}_{jl}-\partial_{l} \Gamma^{i}_{jk} + \Gamma^{m}_{jl} \Gamma^{i}_{mk} - \Gamma^{m}_{jk} \Gamma^{i}_{ml}. \end{equation}
Contracting indices gives the Ricci tensor $R_{ij}$ and the Ricci scalar $R$,
\begin{equation} R_{ij} = R^{l}{}_{ilj} \ , \ R = g^{ij} R_{ij} \ , \end{equation}
which are useful for determining the curvature content of the manifold $ M $.
Geodesics are defined in local coordinates by
\begin{align}\label{geo} \frac{d^2 \lambda^{i}}{d \tau^2} + \Gamma^{i}_{jk} \frac{d \lambda^j}{d \tau} \frac{d \lambda^k}{d \tau} = 0 \ . \end{align}

%\mrd{\section{Results}
\section{Optimal protocols}
Though one can write down the geodesic equations for the metric Eq.~\eqref{betafric} in the $(y_{0},\beta,k)$ coordinate system, more insight is gained by finding a suitable change of coordinates. Consider the lower right $ 2 \times 2 $ block of the metric Eq.~\eqref{betafric} which is the metric tensor for the two-dimensional $ (\beta,k) $ submanifold. A direct calculation of this submanifold's Ricci scalar yields $ R = -2 \zeta^{c} / m$ which is constant and always strictly negative.

Theorems from Riemannian geometry~\cite{doCarmo} imply that this constant negative-curvature submanifold is isometrically related to the hyperbolic plane. In our construction, we choose the Poincar\'{e} half-plane representation of the hyperbolic plane, which is described by $ \{ (z,x) \in \mathbb{R}^{2} , x > 0 \} $ with metric tensor given by the line element $ ds^2 = g_{ij} dx^{i} dx^{j} =  \frac{dx^2+dz^2}{x^2}  $. The geodesics of the hyperbolic plane (see Fig.~\ref{fig:geofig}) are half-circles with centers on the $ z $-axis and lines perpendicular to the $ z$-axis. Fig.~\ref{fig:geofig}(c) shows two geodesics in $ (z,x) $ coordinates. The portion of the hyperbolic plane $ \{ (z,x) \in \mathbb{R}^{2} , x > 1 , z \in [0, \pi] \} $ may be isometrically embedded in $ \mathbb{R}^{3} $ using the map
\begin{equation} \left( \frac{1}{x} \cos{z} , \log{ \left(\sqrt{x^2-1}+x \right) }-\frac{\sqrt{x^2-1}}{x}, \frac{1}{x} \sin{z} \right). \end{equation}  The geodesics of Fig.~\ref{fig:geofig}~(c) and the part of the hyperbolic plane containing them are embedded in $ \mathbb{R}^3 $ in Fig.~\ref{fig:geofig}(d).

The line element associated with the submanifold metric tensor,
\begin{eqnarray}\label{line} ds^{2} & = & \frac{m}{4 \zeta^{c}} \Bigg[  \frac{1}{\beta^2} \left(4 + \frac{\left( \zeta^{c}\right)^2}{k m}  \right) d \beta^2 + \\ && \frac{2}{\beta k} \left( 2 + \frac{\left( \zeta^{c}\right)^2}{k m} \right) d\beta \, d k + \frac{1}{k^2 }\left( 1+ \frac{\left( \zeta^{c}\right)^2}{k m} \right) dk^2 \Bigg] \nonumber \ , \end{eqnarray}
is coordinate-invariant since it measures geometric distances. Thus we may construct an explicit coordinate transformation,
\begin{equation}
\label{confcoord} x \equiv \frac{1}{2 \beta \zeta^{c}} \sqrt{\frac{m}{k}} \ , \ z \equiv \frac{1}{4 \beta k} \ ,
\end{equation}
to demonstrate the equivalence of the submanifold with a portion of the Poincar\'{e} plane. Note that $ x $ is proportional to the classical partition function of the system in equilibrium, and $ z $ is proportional to the equilibrium variance of $ y-y_{0} $. Inverting Eq.~\eqref{confcoord}, and substituting into Eq.~\eqref{line} gives the metric tensor in $(z,x)$-coordinates,
\begin{equation}\label{confmetric} ds^2 = \frac{m}{\zeta^{c}} \frac{ dx^2 + dz^2}{x^2} \ . \end{equation}

The line element corresponding to the metric of the full three-dimensional manifold in Eq.~\eqref{betafric} is
\begin{equation}
\label{3Dmetric}
ds^2 = \frac{m}{\zeta^c} \frac{z \, dy_{0}^2 + dx^2 + dz^2}{x^2}
\end{equation}
in $ (y_{0}, z, x ) $ coordinates.
To fully exploit the machinery of Riemannian geometry to find closed-form geodesics, we look for Killing fields of Eq.~\eqref{3Dmetric}. In general~\cite{Jost,doCarmo,Carroll}, isometries of a metric are generated by the Killing vector fields $ K $ which are themselves characterized by the Killing equation
\begin{equation}\label{Killing} \nabla_{i} K_{j} + \nabla_{j} K_{i} = \partial_{i} K_{j} + \partial_{j} K_{i} - 2 \Gamma^{k}_{ij} K_{k} = 0 \ . \end{equation}
While directly solving this system of equations may be difficult, certain characterizations of Killing vectors help circumvent this difficulty. For instance, if in a given coordinate system the metric tensor components are independent of a coordinate $ x^{i} $, then the coordinate vector $ \partial_{x^{i}} $ is a Killing field~\cite{Carroll}. Hence, $ \partial_{y_{0}} $ is clearly a Killing vector field. Examining the full set of Killing equations shows that
\begin{equation} K = y_{0} \partial_{y_{0}} + 2 x \partial_{x} + 2 z \partial_{z} \end{equation}
is also a Killing vector field. There may be more solutions to the Killing equation yet to be discovered.

In general~\cite{Carroll}, for Killing vector $K_i$ the quantity $ K_{i} \frac{d \lambda^{i}}{d \tau} $ is conserved along the geodesic described by $ \boldsymbol \lambda $.
This follows from
\begin{equation} \frac{d}{d \tau } \left(  K_{i} \frac{d \lambda^{i}}{d \tau} \right) = \nabla_{i} K_{j} \frac{d \lambda^{i}}{d \tau} \frac{d \lambda^{j}}{d \tau} + K_{i} \frac{D}{d \tau} \frac{d \lambda^{i}}{d \tau} = 0. \end{equation}
The first term of the equation vanishes by the definition of the Killing field and the second term vanishes by the geodesic equation Eq.~\eqref{geo}. For the three-dimensional inverse diffusion tensor, we have the following two conserved quantities associated with Killing fields:
\begin{equation}\label{conserved} \frac{z(\tau)}{x^2(\tau)} \frac{d y_{0}}{d \tau} \ , \ \frac{2}{x(\tau)} \frac{d x}{d \tau} + \frac{z(\tau)}{x^2(\tau)} y_{0}(\tau) \frac{dy_{0}}{d \tau} + \frac{2 z(\tau)}{x^2(\tau)} \frac{d z}{d \tau}. \end{equation}
%Since $ K = \partial_{y_{0}} $ is a Killing vector field, any geodesic must satisfy
%\begin{equation}\label{Killingimp} \frac{z(\tau)}{x^2(\tau)} \frac{d y_{0}}{d \tau} = \mbox{Constant} \ . \end{equation} (See Supplementary Material for relevant background on Killing fields and conserved quantities.)

To solve the geodesic equations, note that the velocity of the geodesic (\emph{i.e.}, its tangent vector) must have constant norm~\cite{Jost, doCarmo,Carroll}. For convenience, we choose the norm so that
\begin{equation}\label{speed} 1 = \frac{1}{x^2(\tau)} \left( \left( \frac{d z}{d \tau}  \right)^2 + \left( \frac{d x}{d \tau}  \right)^2\right) + \frac{c_{1}^2}{r^2} \frac{x^2(\tau)}{z(\tau)} \  \end{equation}
where we have used the first conserved quantity of Eq.~\eqref{conserved}. We combine this with the full geodesic equation for $ x(\tau)$, to decouple $ x(\tau) $ from $ y_{0}(\tau) $ and $ z(\tau) $:
\begin{equation} \frac{d^2 x}{d \tau^2} - \frac{2}{x(\tau)} \left( \frac{d x}{d \tau}\right)^2 + x(\tau) =0 \ , \end{equation} which has solution
\begin{equation}\label{3dx} x(\tau) = r \operatorname{sech}(\tau) \ . \end{equation}
When $ z(\tau) $ is constant, the geodesic equation for $ z $ implies that $ y_{0} $ is also constant, giving a geodesic straight line in the constant-$z$ submanifold.

When $ z(\tau) $ is not constant, Eqs.~\eqref{speed} and~\eqref{3dx} imply
\begin{equation} \frac{x^4(\tau)}{r^2} = \left( \frac{d z}{d\tau} \right)^2+\frac{c_{1}^2}{r^2} \frac{x^4(\tau)}{z(\tau)} \ , \end{equation}
which integrates to
\begin{equation} z(\tau)= h^{-1}\big(c_{2}- r \tanh(\tau) \big), \end{equation}
where
\begin{equation}
h \left(\xi \right)  \equiv  \xi \sqrt{1-\tfrac{c_{1}^2 }{\xi}} + \tfrac{1}{2} c_{1}^2 \log \Big(    2 \xi \big( 1+\sqrt{ 1-\tfrac{c_{1}^2}{\xi} } \big) -c_{1}^2 \Big) \ .
\end{equation}
The Killing conserved quantities of Eq.~\eqref{conserved}, together with $ x(\tau) $ and $ z(\tau) $, yield
\begin{eqnarray}\label{3dy0} y_{0}(\tau) & = &   E- c_{1} \log \Bigg[ - c_{1}^2 +  2 h^{-1} \left( c_{2} - r \tanh (\tau) \right) \times \nonumber \\ && \left( 1 + \sqrt{ 1 - \frac{ c_{1}^2}{h^{-1} \left( c_{2} - r \tanh (\tau) \right)}} \right) \Bigg] \ .
\end{eqnarray}

Let $ (y_{0,i} , x_{i},z_{i}) $ and $ (y_{0,f} , x_{f},z_{f}) $ denote the endpoints of the geodesic. Define $ \Delta \lambda \equiv \lambda_f - \lambda_i $ and $ \bar{\lambda} \equiv \frac{\lambda_i+\lambda_f}{2} $ for $ \lambda \in \{y_0,x,z \} $. Defining $ \bar{h} \equiv \frac{h \left(z_{f} \right)+h \left(z_{i} \right) }{2} $ and $ \triangle h \equiv h \left( z_{f} \right)-h\left( z_{i} \right) $ , the constant $ c_{2} $ may be written as
\begin{equation}\label{ceq} c_{2} = \bar{h}+ \bar{x} \frac{ \triangle x  }{\triangle h}  \end{equation}
and $ r $ is given by
\begin{equation} r^2 = x_{i}^2 + \frac{1}{4} \Bigg( \triangle h + 2 \frac{\triangle x}{\triangle h} \bar{x} \Bigg)^2. \end{equation}
The constant $ E $ is given by
\begin{equation} E = y_{0,i} + c_{1} \log \left( -c_{1}^2 + 2 z_{i} \left( 1+ \sqrt{ 1- \frac{c_{1}^2}{z_{i}}} \right) \right) \end{equation}
and $ c_{1} $ is determined by the equation
\begin{eqnarray} \triangle y_{0} & = & - c_{1} \Bigg[ \log \left( -c_{1}^2 + 2 z_{f} \left( 1+ \sqrt{ 1- \frac{c_{1}^2}{z_{f}}} \right) \right)- \nonumber \\ &&   \log \left( -c_{1}^2 + 2 z_{i} \left( 1+ \sqrt{ 1- \frac{c_{1}^2}{z_{i}}} \right) \right) \Bigg] . \end{eqnarray}
The parameter $ \tau $ ranges between the values
 \begin{equation} \tau_{i} = \mbox{sgn} \left(\triangle h + 2 \frac{\triangle x}{\triangle h} \bar{x} \right) \mbox{sech}^{-1} \left( \frac{x_{i}}{r} \right) \end{equation} and \begin{equation}  \tau_{f} = -\mbox{sgn} \left( \triangle h - 2 \frac{\triangle x}{\triangle h} \bar{x} \right) \mbox{sech}^{-1} \left( \frac{x_{f}}{r} \right).   \end{equation}
%The conserved \prz{quantities} of Eq.~\eqref{conserved} may be used to show that $\label{3dx} x(\tau) = r \operatorname{sech}(\tau)$, $\label{3dz} z(\tau)= h^{-1}\big(c_{2}- r \tanh(\tau) \big)$, and
%\begin{eqnarray}\label{3dy0} y_{0}(\tau) & = &   E- c_{1} \log \Bigg[ - c_{1}^2 +  2 h^{-1} \left( c_{2} - r \tanh (\tau) \right) \times \nonumber \\ && \left( 1 + \sqrt{ 1 - \frac{ c_{1}^2}{h^{-1} \left( c_{2} - r \tanh (\tau) \right)}} \right) \Bigg] \
%\end{eqnarray}
%satisfy the geodesic equations, where
%\begin{equation}
%h \left(\xi \right)  \equiv  \xi \sqrt{1-\tfrac{c_{1}^2 }{\xi}} + \tfrac{1}{2} c_{1}^2 \log \Big(    2 \xi \big( 1+\sqrt{ 1-\tfrac{c_{1}^2}{\xi} } \big) -c_{1}^2 \Big) \ .
%\end{equation}

\begin{figure}[b]
\centering
\includegraphics{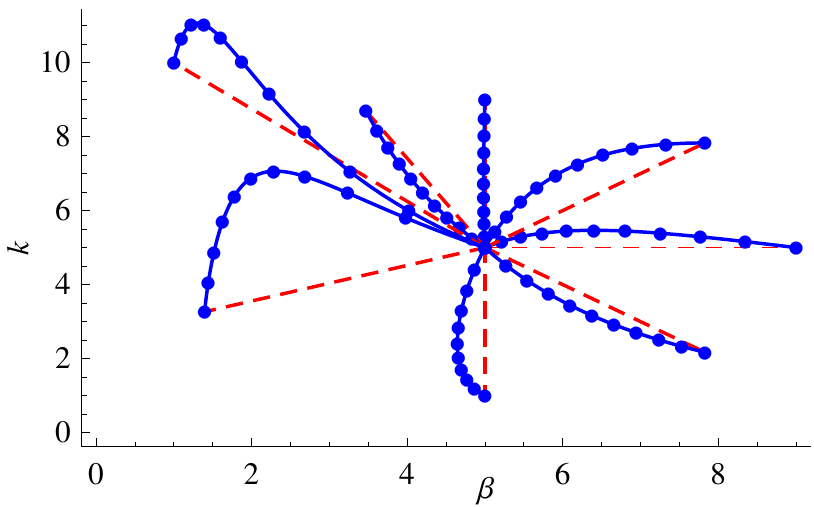}
\caption{Optimal protocols differ substantially from linear interpolation (red dashed lines). Blue solid curves represent geodesics of the inverse diffusion tensor, and are thus optimal protocols for transitioning the system from one state to another in a fixed amount of time. Blue dots indicate points separated by equal times along each of the eight optimal paths shown.}
\label{fig:geodesics}
\end{figure}

%
%Finally, Eq.~\eqref{Killingimp} may be integrated to obtain
When $ y_{0} $ is held fixed, the geodesics are precisely those of the hyperbolic plane as expected. Furthermore, these geodesics are necessarily minimizing by virtue of the constant, negative Ricci scalar~\cite{Jost,doCarmo}. Several example geodesics are displayed in Fig.~\ref{fig:geodesics}.
\\
\section{Computing dissipation numerically}
We validate the optimality of these geodesics by calculating excess work directly from the Fokker-Planck equation. In full generality, the mean excess work as a functional of the protocol $ \boldsymbol \lambda (t) = (y_{0}(t),\beta(t),k(t)) $ is
\begin{subequations}
\begin{align} \left< \left( \beta W \right)_{\rm ex} \right>  &\equiv \int_{0}^{t_{f}} dt \ \beta  \mathcal{P}_{\rm ex}  \\
&= \int_{0}^{t_{f}} dt \Bigg( \dot{\beta} \frac{ \langle p^2 \rangle}{2m} + \frac{1}{2} \langle \left(y -y_{0} \right)^2 \rangle \left( k \dot{\beta} + \dot{k} \beta \right) \nonumber \\ & \ \ \ \ \ + k \beta \dot{y}_{0} \langle y_{0}-y \rangle - \frac{\dot{\beta}}{ \beta}- \frac{ \dot{k}}{2 k } \Bigg) \ . \label{funcexwork} \end{align} \end{subequations}
Here angled brackets denote averages over the nonequilibrium probability density $ f(y,p,t) $.

\begin{figure}[t]
\centering
\includegraphics{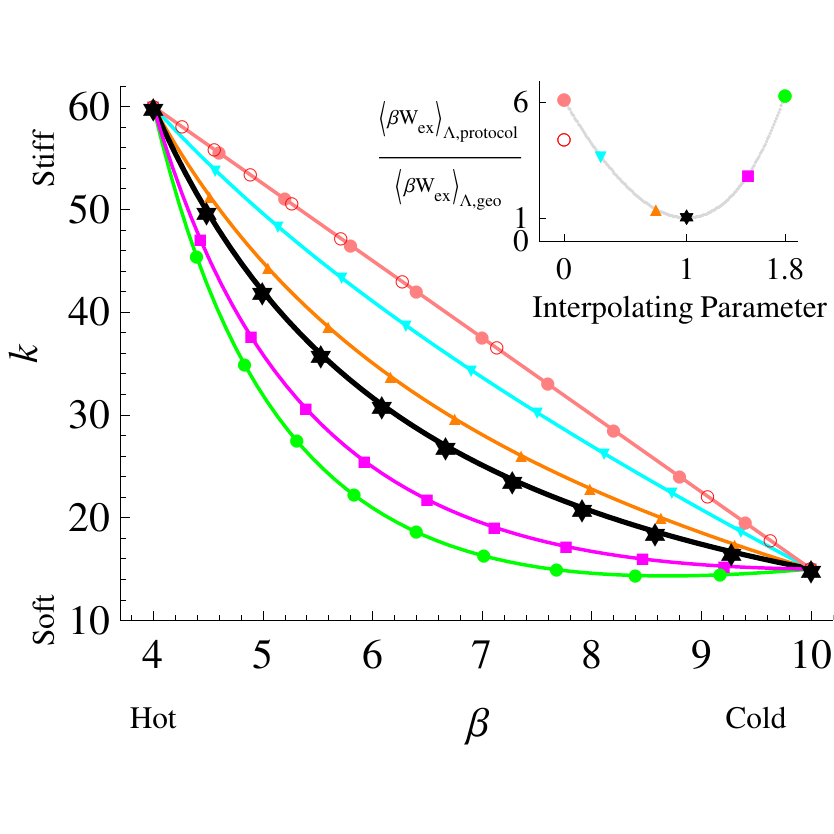}
\caption{Geodesics describe protocols that outperform naive straight line paths in parameter space. A geodesic between two fixed points in the $ (\beta,k)$-plane (black) and several comparison protocols are pictured above. The comparison protocols were generated via a linear interpolation between the constant speed straight line (pink) and the geodesic. The tick marks represent points separated by equal times. The solid pink dots correspond to the constant speed parametrization of the line whereas the open red circles correspond to the optimal parametrization along this straight path.
The ratio of excess work to that of the geodesic protocol is: $ 6.12 $ (pink circle), $ 4.37 $ (red open circle), $ 3.67 $ (cyan downward-pointing triangle), $ 1.38 $ (orange upward-pointing triangle), $ 1.00 $ (black star (geodesic)), $ 2.86 $ (magenta square), and $ 6.29 $ (green circle). These ratios are plotted in the inset figure along with a graph of the ratio as a function of the interpolating parameter (light gray curve).
All excess work values were calculated using the Fokker-Planck system Eq.~\eqref{FPsys}. Here, $A=10^{-2}, B=10^{-3}, M=1$, placing the system within the near-equilibrium regime and ensuring accuracy of the inverse diffusion tensor approximation.}
\label{fig:compare}
\end{figure}

Standard arguments~\cite{Zwanzig2001} yield the Fokker-Planck equation for the time evolution of $ f(y,p,t) $,
\begin{equation}\label{FokkerPlanck} \frac{\partial f}{\partial t} + \frac{p}{m} \frac{\partial f}{\partial y} - k(t) \left[ y -y_{0}(t) \right] \frac{\partial f}{\partial p} - \frac{\zeta^{c}}{m} \frac{\partial [ p f]  }{\partial p} -  \frac{ \zeta^c}{\beta(t)} \frac{\partial^2 f}{\partial p^2} =0 \ . \end{equation}
By integrating Eq.~\eqref{FokkerPlanck} against $ y, p, $ etc., we find a system of equations for relevant nonequilibrium averages:
 \begin{subequations}
 \label{FPsys}
\begin{align}
 \frac{d}{dt} \langle y \rangle   &=  \frac{\langle p \rangle}{m} \\
 \frac{d}{dt} \langle p \rangle   &= -\frac{\zeta^{c}}{m} \langle p \rangle - k \langle y-y_{0} \rangle \\
 \frac{d}{dt} \langle p y \rangle    &= \frac{\langle p^2 \rangle}{m} -k \langle y^2 \rangle - \frac{\zeta^c}{m} \langle p y \rangle + k y_{0} \langle y \rangle  \\
 \frac{d}{dt} \langle y^2 \rangle  &= \frac{2}{m}  \langle p y \rangle \\
 \frac{d}{dt} \langle p^2 \rangle   &= -2 k \langle p \left( y -y_{0} \right)\rangle -2 \frac{\zeta^{c}}{m} \langle p^2 \rangle + \frac{2 \zeta^c}{\beta}. \
 \end{align}
\end{subequations}
Following the derivation of the friction tensor in~\cite{Sivak2012b} would require us to use linear response theory and to supplement the system Eq.~\eqref{FPsys} by initial conditions
\begin{subequations}
\label{FPsysinitconds}
\begin{align}
\left< y \right>(0) &= y_{0}(0) \\
\left< p \right>(0) &= 0 \\
\left< y^2 \right>(0) &= y_{0}(0)^2 + \frac{1}{k(0) \beta(0) }  \\
\left< py \right>(0) &= 0 \\
\left< p^2 \right>(0) &= \frac{m}{\beta(0)} \ .
\end{align}
\end{subequations} We solve these equations numerically and compare a geodesic protocol with naive protocols in Fig.~\ref{fig:compare}.

This system has three natural dimensionless quantities
\begin{equation}
\label{eq:dim_params}
A \equiv \frac{m}{\zeta^{c} \triangle t} \ , \ B \equiv \frac{\zeta^{c}}{ \tilde{k} \triangle t} \ , \ M \equiv \frac{\zeta^{c} \left( \triangle t \right)^3}{\tilde{l}^2 m^2 \tilde{\beta}} \ ,
%\Gamma \equiv \frac{\tilde{k} \left( \triangle t\right)^2}{m} \ , \ Z \equiv \frac{\zeta^{c} \triangle t}{m} \ , \ M \equiv \frac{\zeta^{c} \left( \triangle t \right)^3}{\tilde{l}^2 m^2 \tilde{b}} \ ,
\end{equation}
dependent upon characteristic scales for (inverse) temperature $\tilde{\beta}$, length $\tilde{l}$, spring constant $\tilde{k}$ and the protocol duration $\triangle t$.
These suggest at least two plausible measures of distance from equilibrium~\cite{Sivak2012b}. $A$ corresponds to the ratio of two timescales, the timescale $\frac{m}{\zeta^c}$ for frictional damping and the timescale of the perturbation protocol $\Delta t$. Likewise, $B$ is the ratio of two powers during changes of $y_0$, the dissipative power $\zeta^c (\Delta y_0/\Delta t)^2$ and the elastic power $\tilde{k}(\Delta y_0)^2 / \Delta t$. As $A$ decreases and as $B$ decreases, the system will remain closer to equilibrium during the course of the nonequilibrium perturbation, and hence our near-equilibrium approximation will be more accurate.
%Specifically, the ratio $\frac{m}{\zeta^c}$ represents a characteristic timescale of dissipation due to external forcing of the system; if this is small compared to the protocol duration $\Delta t$, then we might expect the system to remain close to equilibrium.
%Moreover, the limit of large $Z$ corresponds to overdamped (Brownian) motion. Specifically, the ratio $\frac{m}{\zeta^c}$ represents a characteristic timescale of dissipation due to external forcing of the system; if this is small compared to the protocol duration $\Delta t$, then we might expect the system to remain close to equilibrium.
%Increasing $\triangle t$ (with all other scales and constants fixed) slows the driving rate while maintaining the relaxation rate, and thus should improve the approximation.

This intuition is confirmed in our numerical calculations: with $A \ll 1$ and $B \ll 1$, the dissipation of geodesic protocols obtained numerically via Fokker-Planck agrees with the inverse diffusion tensor approximation to better than $.1\%$ (see Fig.~\ref{fig:compare}). Note that, while the inverse diffusion tensor approximation is excellent for optimal protocols and small deviations thereof, it can deviate substantially from the exact result for large deviations from the geodesic.
\\
\section{The inverse diffusion tensor arises naturally from the Fokker-Planck equation}
 If we neglect terms involving derivatives of protocols of degree two and higher, we may find an approximate solution to the Fokker-Planck system:
\begin{subequations}
\begin{align}
\left< y \right> &\approx y_{0} -\frac{\zeta^c}{k} \dot{y}_{0}  \\
\left< p \right> &\approx m \dot{y}_{0} \\
\left< p y \right> &\approx m y_{0} \dot{y}_{0} - \frac{m}{2} \left( \frac{\dot{k}}{\beta k^2} + \frac{\dot{\beta}}{\beta^2 k}  \right) \\
\left< p^2 \right> &\approx \frac{m}{\beta} + \frac{m^2}{\zeta^c} \left( \frac{\dot{k}}{2 \beta k} + \frac{\dot{\beta}}{\beta^2 } \right) \\
\left< y^2 \right> &\approx y_{0}^2 + \frac{1}{\beta k}-\frac{2 \zeta^{c}}{k} y_{0} \dot{y}_{0} \\
&+ \dot{k} \left( \frac{m}{\zeta^c} \frac{1}{2 \beta k^2} + \frac{\zeta^c}{2 \beta k^3} \right) + \dot{\beta} \left( \frac{m}{\zeta^c} \frac{1}{ \beta^2 k} + \frac{\zeta^c}{2 \beta^2 k^2}  \right). \  \nonumber
\end{align}
\end{subequations} \\
Substituting this into the expression for mean excess power Eq.~\eqref{funcexwork}, we recover Eq.~\eqref{etadef}. The argument above suggests that the emergence of the inverse diffusion tensor from the Fokker-Planck equation may follow from a perturbation expansion in small parameters.\\

\section{Discussion}
We have employed geometric techniques to find optimal protocols for a simple, but previously unsolved, stochastic system. Calculation of the Ricci scalar for a submanifold pointed to a change of coordinates that identified the submanifold with the hyperbolic plane and greatly simplified the metric for the full three-dimensional manifold. This simplification, combined with the identification of a Killing field, permitted calculation of an exact closed-form expression for geodesics. Exact calculations using the Fokker-Planck equation confirmed that geodesics in the $ \left(\beta, k \right)$-submanifold do indeed produce less dissipation than any comparison protocol we tested.

In addition to being useful for identifying optimal protocols, we expect that the Ricci scalar will turn out to have an important physical interpretation. Riemannian geometry has been useful for the study of thermodynamic length of macroscopic systems~\cite{Brody2009, Ruppeiner2010}, and there has been some speculation about the role of the Ricci scalar in that setting~\cite{Ruppeiner2010}, but the interpretation of~$R$ arising from the inverse diffusion tensor remains ambiguous. We hope that further study of these geometrical ideas extended to nonequilibrium systems will help clarify its role.

It would also be interesting to establish a physical interpretation for the conserved quantities arising from Killing fields in this context.
We found two conserved quantities (see Eq.~\eqref{conserved}), which may be the only ones, but this model could have as many as six, given that there might be as many as six unique globally smooth Killing fields for this three-dimensional model system. (In general, there are at most $\frac{1}{2} n \left( n+1 \right) $ independent globally smooth Killing fields where $ n $ is the dimension of the manifold~\cite{Carroll}.)

In the course of developing our framework, we encountered four distinct measures of the departure from equilibrium. The first two were dimensionless parameters, $A$ and $B$, which have relatively straightforward physical interpretations --- the timescale for frictional dissipation relative to the protocol duration and the ratio of the dissipative power to elastic power, respectively (see discussion following Eq.~(\ref{eq:dim_params})).

The third was the disagreement between dissipation computed assuming linear response theory and the true dissipation.  Empirically, we found that our linear response approximation was consistently accurate for all parameter regimes we tested in which both dimensionless parameters $A$ and $B$ were small, at least for protocols not too far from geodesics.  Conversely, the linear response approximation appeared to break down for many cases we tested with at least one of these parameters of order unity or greater.  However, the full extent of validity of the linear response approximation is not clear to us, suggesting an important direction for future research.

Finally, we found that truncating to first order in temporal derivatives of the control parameters in our model was sufficient to yield the same inverse diffusion tensor formalism we originally derived using linear response theory. While it is plausible that these two types of linear approximations are directly related, further exploration is needed to uncover the relationship between linear response theory and truncating the model equations to first order in temporal derivatives.

Our results are novel in three distinct ways. First, we included $\beta$ as a control parameter, which is a natural extension of thermodynamic length ({\em e.g.}~\cite{Shenfeld,Brody2009}) that is amenable to direct experimental confirmation. Our work generalizes the construction of~\cite{Sivak2012b} and opens up new experimental avenues for testing the validity of the framework.

%Secondly, our geodesic protocols optimize dissipation for simultaneous variation of all three adjustable parameters.
Secondly, our geodesic protocols optimize dissipation for simultaneous variation of all three adjustable parameters; to our knowledge, no previous study has reported optimal protocols for any model system with three control parameters. In~\cite{Seifert2008, Seifert2007}, Seifert and coworkers elegantly derived the exact optimal protocols for perturbing the position~$ y_{0}$ and spring constant $k$ separately, for both over-damped and under-damped Langevin dynamics. In~\cite{Aurell}, Aurell and coworkers discuss the simultaneous variation of the stiffness and the location of the trap. We note that our method misses the protocol jumps found in their analysis due to our smoothness assumptions on the protocols. When this restriction on the differentiability of the curve is imposed, we found that any component of the optimal protocol~$\big(y_{0}(t), \beta(t), k(t) \big)$ generically depends on all components of both endpoints due to the non-trivial geometry of the parameter space.

Finally, we successfully brought the machinery of Riemannian geometry to bear on a small-scale, nonequilibrium thermodynamic problem, revealing a surprisingly rich geometric structure. Concepts such as Killing vector fields, coordinate invariance and the Ricci scalar proved indispensable in the construction of optimal protocols.
These results are encouraging and this approach may prove useful for understanding the constraints on the non-equilibrium thermodynamic efficiency of biological and synthetic molecular machines.
%\mrd{We are encouraged by these results and we hope that this approach will prove useful for understanding the constraints on the non-equilibrium thermodynamic efficiency of biological and synthetic molecular machines.}
\\
%\mrd{\section{Acknowledgments}}
\acknowledgments
PRZ and MRD would like to thank Tony Bell for many useful discussions and Peter Battaglino for helpful discussions and sharing computer code.  MRD would also like to thank Badr Albanna, Susanna Still, and Jascha Sohl-Dickstein for many valuable discussions. MRD gratefully acknowledges support from the McKnight Foundation, the Hellman Family Faculty Fund, the McDonnell Foundation, and the Mary Elizabeth Rennie Endowment for Epilepsy Research. MRD and PRZ were partly supported by the National Science Foundation
through Grant No. IIS-1219199. DAS and GEC were funded by the Office of Basic Energy Sciences of the U.S. Department of Energy under Contract No. DE-AC02-05CH11231.

\bibliography{geocontrol}

\end{document}